\documentstyle[epsf]{mn}

\newif\ifAMStwofonts

\newcommand{\gtorder}{\mathrel{\raise.3ex\hbox{$>$}\mkern-14mu
                \lower0.6ex\hbox{$\sim$}}}
\newcommand{\ltorder}{\mathrel{\raise.3ex\hbox{$<$}\mkern-14mu
                \lower0.6ex\hbox{$\sim$}}}
\newcommand{\mav}{\mbox{$\left < M \right >$}}
\newcommand{\ovii}{O\,{\sc vii}}
\newcommand{\oviii}{O\,{\sc viii}} 

\newcommand{\heii}{He\,{\sc ii}}  
\newcommand{\civ}{C\,{\sc iv}} 
\newcommand{\apj}{ApJ}
\newcommand{\apjs}{ApJS}
\newcommand{\aap}{A\&A}
\newcommand{\mnras}{MNRAS}
\newcommand{\pasj}{PASJ}

\ifoldfss
  \newcommand{\rmn}[1] {{\rm #1}}

  \ifCUPmtlplainloaded \else
    \NewTextAlphabet{textbfit} {cmbxti10} {}
    \NewTextAlphabet{textbfss} {cmssbx10} {}
    \NewMathAlphabet{mathbfit} {cmbxti10} {} 
    \NewMathAlphabet{mathbfss} {cmssbx10} {} 
  \fi
  \ifAMStwofonts
    \ifCUPmtlplainloaded \else
      \NewSymbolFont{upmath} {eurm10}
      \NewSymbolFont{AMSa} {msam10}
      \NewMathSymbol{\upi}     {0}{upmath}{19}
      \NewMathSymbol{\umu}     {0}{upmath}{16}
      \NewMathSymbol{\upartial}{0}{upmath}{40}
      \NewMathSymbol{\leqslant}{3}{AMSa}{36}
      \NewMathSymbol{\geqslant}{3}{AMSa}{3E}

      \let\leq=\leqslant \let\le=\leqslant
      \let\geq=\geqslant 
    \fi
  \fi
\fi 

\ifnfssone
  \newmathalphabet{\mathit}
  \addtoversion{normal}{\mathit}{cmr}{m}{it}
  \addtoversion{bold}{\mathit}{cmr}{bx}{it}
  \newcommand{\rmn}[1] {\mathrm{#1}}

  \newmathalphabet{\mathbfit} 
  \addtoversion{normal}{\mathbfit}{cmr}{bx}{it}
  \addtoversion{bold}{\mathbfit}{cmr}{bx}{it}
  \newmathalphabet{\mathbfss} 
  \addtoversion{normal}{\mathbfss}{cmss}{bx}{n}
  \addtoversion{bold}{\mathbfss}{cmss}{bx}{n}
  \ifAMStwofonts
    \ifCUPmtlplainloaded \else
      \UseAMStwoboldmath
      \makeatletter
      \new@mathgroup\upmath@group
      \define@mathgroup\mv@normal\upmath@group{eur}{m}{n}
      \define@mathgroup\mv@bold\upmath@group{eur}{b}{n}
      \edef\UPM{\hexnumber\upmath@group}
      \new@mathgroup\amsa@group
      \define@mathgroup\mv@normal\amsa@group{msa}{m}{n}
      \define@mathgroup\mv@bold\amsa@group{msa}{m}{n}
      \edef\AMSa{\hexnumber\amsa@group}
      \makeatother
      \mathchardef\upi="0\UPM19
      \mathchardef\umu="0\UPM16
      \mathchardef\upartial="0\UPM40
      \mathchardef\leqslant="3\AMSa36
      \mathchardef\geqslant="3\AMSa3E

      \let\leq=\leqslant \let\le=\leqslant
      \let\geq=\geqslant 
    \fi
  \fi
\fi 

\ifnfsstwo
  \newcommand{\rmn}[1] {\mathrm{#1}}

  \DeclareMathAlphabet{\mathbfit}{OT1}{cmr}{bx}{it}
  \SetMathAlphabet\mathbfit{bold}{OT1}{cmr}{bx}{it}
  \DeclareMathAlphabet{\mathbfss}{OT1}{cmss}{bx}{n}
  \SetMathAlphabet\mathbfss{bold}{OT1}{cmss}{bx}{n}
  \ifAMStwofonts
    \ifCUPmtlplainloaded \else
      \DeclareSymbolFont{UPM}{U}{eur}{m}{n}
      \SetSymbolFont{UPM}{bold}{U}{eur}{b}{n}
      \DeclareSymbolFont{AMSa}{U}{msa}{m}{n}
      \DeclareMathSymbol{\upi}{0}{UPM}{"19}
      \DeclareMathSymbol{\umu}{0}{UPM}{"16}
      \DeclareMathSymbol{\upartial}{0}{UPM}{"40}
      \DeclareMathSymbol{\leqslant}{3}{AMSa}{"36}
      \DeclareMathSymbol{\geqslant}{3}{AMSa}{"3E}

      \let\leq=\leqslant \let\le=\leqslant
      \let\geq=\geqslant 
    \fi
  \fi
\fi 

\ifCUPmtlplainloaded \else
  \ifAMStwofonts \else 
    \def\upi{\pi}
    \def\umu{\mu}
    \def\upartial{\partial}
  \fi
\fi

\title{Radiation Pressure Acceleration by X-rays in Active Galactic Nuclei}
\author[Doron Chelouche and Hagai Netzer]
       {Doron Chelouche\thanks{email: doron@wise.tau.ac.il; netzer@wise.tau.ac.il} and Hagai
         Netzer\mbox{\raise.9ex\hbox{$\star$}} \\
        School of Physics and Astronomy and the Wise Observatory,
        The Beverly and Raymond Sackler Faculty of Exact Sciences,\\
        Tel Aviv University, Tel Aviv 69978, Israel}

\pubyear{2000}

\begin{document}
\maketitle

\label{firstpage}

\begin{abstract}
We present  calculations of the dynamics of highly
ionized gas clouds that are confined by  external pressure, and 
are ionized by  AGN continuum. We focus on the gas that is seen in absorption in
the X-ray spectrum of many AGN and show that  
 such gas can reach hydrostatic equilibrium under various conditions. The
principal conclusion is that the clouds can be accelerated to high velocities by
 the central X-ray source. The
dynamical problem can be reduced to the calculation of a single
parameter, the average force multiplier, $\left< M \right >$. The typical value
 of $\left< M \right >$ is  $\sim 10$ suggesting that radiation
 pressure acceleration by X-rays is efficient for $L/L_{\rmn{Eddington}}
 \gtorder 0.1$. 
The terminal velocity scales with the escape velocity at the
  base of the flow 
and can exceed it by a large factor.
The typical velocity for a HIG flow that originates at $R=10^{17}~\rmn{cm}$ in 
a source with $L_x=10^{44}$ $\rmn{erg~s}^{-1}$  is
$\sim1000~\rmn{km~s}^{-1}$, i.e. similar to 
the velocities observed in several X-ray and UV absorption systems.

Highly ionized AGN clouds are driven mainly by
bound-free absorption and 
bound-bound processes are less important unless the lines are
significantly broadened or the column density is very small. Pressure
laws that result in constant or  
outward decreasing ionization parameters are most effective in
accelerating the  flow. 
\end{abstract}

\begin{keywords}
ISM: jets and outflows ---
galaxies: active ---
galaxies: nuclei ---
quasars: absorption lines ---
X-rays: general
\end{keywords}

\section{Introduction}
Highly ionized gas (HIG) is common in both types of active galactic
nuclei (AGN). It is seen as strong absorption features in many type-I (Seyfert
1) galaxies (Reynolds 1997; George et al. 1998, Kaspi et al. 2000 and
references therein) and as strong, large equivalent width emission
lines in many type-II (Seyfert 2) AGN (e.g. 
Turner et al. 1997; Netzer, Turner \& George 1998). The situation regarding the
high luminosity AGN (the bright quasars, see George et al. 2000 for review and
references) is still unclear, mostly because of the limited
signal-to-noise ASCA observations used in such studies.

The strongest observed HIG features in type-I AGN are several
absorption edges, at around 
1 keV, mostly due to \ovii\ (0.74keV) \& \oviii\ (0.87keV). The deduced hydrogen
column density  is in the range of $10^{21}-10^{24}$ cm$^{-2}$,
the covering fraction is close to unity and the
ionization parameter about two orders of magnitude larger than the broad line region
(BLR) ionization parameter (George et al. 1998). Recent {\it Chandra}
observations reveal the presence of numerous absorption and emission
lines (e.g., Kaspi et al. 2000). The absorption lines are narrow
($\sim 150~{\rm km~s^{-1}}$)and
exhibit outflow velocities of $\ltorder 1000~\rmn{km~s}^{-1}$. Currently, there
is no clear indication regarding the 
location and hence the mass of this component. The very few observations of variable
absorption features (e.g., Guainazzi et al. 1996; and references therein)
suggest a dimension not too different from that 
of the BLR. If correct, this would mean that the HIG observed in type-II AGN  is of
different origin since its distance from the centre must be larger than $\sim 10$ pc.
 
Many type-I AGN  show also narrow  UV absorption features due to  
  \civ $\lambda1548$ and other resonance lines. In most cases the lines are
 blueshifted with respect to the systemic velocity, implying outflow 
 with typical velocities
of $\sim 1000~\rmn{km~s}^{-1}$. In several well studied cases the lines are
resolved, showing intrinsic 
width of $100-300~\rmn{km~s}^{-1}$ (Crenshaw et al. 1999; Crenshaw \& Kraemer
1999; Srianand 2000;  and references
therein). There have been several suggestions linking the origin and dynamics of the
UV and X-ray absorption features (Mathur et al. 1994;
Mathur et al. 1995; Murray \& Chiang 1995). This issue is still under discussion (e.g.
the new observations of 3C288.1 by Hamann, Netzer \& Shields 2000).

Radiation pressure driven
flows are thought to be common in astrophysical environments.  
 Stellar winds can be accelerated to their observed velocities
($\gtorder 1000~\rmn{km~s}^{-1}$) by radiation pressure force
(e.g., Castor, Abbott \& Klein 1975). Broad
absorption line (BAL) flows in AGN (BALQSO) can be driven to sub-relativistic
velocities ($\sim 30,000~\rmn{km~s}^{-1}$) by the same physical mechanism
(e.g., Arav, Li \& Begelman 1994; Murray et al. 1995; Proga, Stone \& Kallman
2000)  although other mechanisms have also been proposed (Begelman, de Kool
\& Sikora 1991). In both cases,
 the absorption line widths are of the order of
the outflow velocity which lead to a continuous flow geometry. In this
case, the
physics can be described by the Sobolev
approximation which is valid provided the velocity changes on length
scales over which the other gas properties (e.g., ionization
state, temperature, density) do not change considerably.
Obviously not all types of flow
can be driven by radiation pressure forces, among these are jets and
supernovae envelopes.

A second, almost orthogonal approach to the study of flows
 is to consider moving ``clouds'' or condensations (hereafter, the
 ``cloud model''). Such clouds are
 confined by means of external pressure and are likely to be in hydrostatic
equilibrium.  These models have been studied thoroughly in the
context of broad line region (BLR) dynamics (e.g., Blumenthal \&
Mathews 1979; Mathews 
1986). The clouds are assumed to be accelerated ballistically, with
no appreciable inner velocity gradients and their spectra consists
 of narrow, blueshifted absorption lines (e.g.,
Mathews 1975). Cloud confinement and stability 
 have been studied, extensively, in many
works (e.g., Krolik 1979; Mathews 1986; Mathews 1992).
Possible confinement mechanisms include thermal confinement
(e.g., Krolik, McKee \& 
Tarter 1981) with its well known problems 
(e.g., Mathews \& Ferland 1987),
and nonthermal confinement, e.g., by magnetic fields  (Rees
1987; Bottorff, Korista \& Shlosman 2000). Despite the many
unresolved issues, the cloud
model has had great success in accounting for the observed  features
of the BLR clouds (e.g. Netzer 1990), NLR clouds  (e.g., Kraemer, Ruiz
\& Crenshaw 1998), and  BAL flows (e.g. Arav et al. 1994). 

While acceleration by UV radiation, in AGN, has been studied
extensively, there are only very few works discussing the acceleration by X-ray
radiation (e.g., Mathews \& Capriotti 1985; Mathews \& Veilleux 1989, none in
great detail). In particular,
 there is no systematic study of the 
acceleration of gas with properties similar to those commonly
observed in the X-ray spectrum of many AGN (for a heuristic 
treatment of the problem see Reynolds \& Fabian 1995, 
for a magneto-hydrodynamical model with comparison to
observations see Bottorff, Korista \& Shlosman 2000). As shown below,
considerations of the relevant  
parameter space  within the framework of the cloud model, result
in very different dynamics for the HIG clouds compared with  previous
works. 

This paper investigates the acceleration of the HIG clouds. 
The purpose is to study the micro-physics of the gas and the radiation
pressure force, and 
to derive the gas velocity under various conditions prevailing in type
I AGN. This will enable to explain observed
spectral features in new, high resolution X-ray spectra.

The paper is organized as follows: in \S2 we outline the 
main ingredients of  the model. Section 3
 describes the calculation procedure, and  present the  results 
concerning  the cloud structure,  the radiation pressure force and the gas
motion, with a following  discussion on the model limitations. The
main conclusions are summarized in \S 4. 

\section{The model}
 
\subsection{Basic assumptions}

Our purpose is to solve for the structure and the dynamics  of HIG clouds in
AGN. Such gas has been detected by ASCA and earlier missions but the
  evidence used here 
 is found mainly in the   recent {\it Chandra} observations such as
the one  reported in Kaspi et al. (2000) for NGC~3783  and in Kaastra et al
(2000) for NGC~5548. In both cases there is a rich absorption spectrum
indicating high level of ionization, large column density ($\sim 10^{22}~{\rm
cm^{-2}}$) and outflow motion  of $\ltorder 1000~{\rm km~s^{-1}}$. The lines
are barely resolved (Kaspi
et al. 2001) and the typical Doppler width is 
($\sim 300~{\rm km~s^{-1}}$). The  observed properties
are, therefore, different from those found in stellar winds and in  BAL
quasars. They can be interpreted as either due to cloud motion or continuous
wind. This paper addresses the first possibility and the wind solution will be
discussed in a future publication.

The main assumptions regarding the cloud model are 
(see also \S3.5):
\begin{enumerate}
\item
The general AGN environment can support various type of clouds. Known
examples
are the BLR and the NLR clouds (e.g. Netzer 1990; see a recent study by Kaiser et al. 2000). 
Such entities with higher level of ionization cannot be ruled out by any known
AGN property and have, in fact, been addressed, by several recent publications
 (e.g., Contini \& Viegas 1999; 
Bottorff et al. 2000).  
The HIG region is modeled as a
 spherical cloud system of large covering factor. The justification is that
such regions are very common in low luminosity AGN (e.g. George et al. 1998).
We concentrate on thin spherical shells although this is not a requirement of
the model (see  \S3.5.)
\item
We assume that the clouds are confined by external pressure $P_{\rm ext}$.
This
is required  since
clouds which are not self-gravitating evaporate on sound crossing
time scales (which is typically much shorter than dynamical time
scales). While the origin of this pressure is not the topic of the
paper, we note that confinement by a hot gas (e.g. Krolik, McKee \&
Tarter 1981) is problematic (Mathews \& Ferland 1987) and magnetic pressure is
the most likely explanation (see Rees (1987) for a general formulation,   de
Kool \& Begelman (1995) for a discussion of radiatively driven, magnetically
confined flows and  Sivron \& Tsuruta (1993) for a discussion of the
magnetically confined cloud model).
 The  pressure is assumed to be time-independent
and its radial dependence is given by   
\begin{equation}
P_{\rmn{ext}} \propto R^{\alpha},
\label{pprof}
\end{equation}
Typically, $-10/3 \leq
\alpha \leq -3/2$ (e.g., Netzer 1990).  The global spherical
symmetry is not critical to our discussion and is relaxed in \S3.5.
\item
We restrict the discussion to geometrically thin HIG clouds, 
$\Delta R/R \ll 1$ where $R$  is the location
of the  {\it illuminated surface} 
 (see for example Reynolds \& Fabian 1995). This is
based on  variability studies which set  rough lower limits on
the density ($\sim 10^6~{\rm cm^{-3}}$, see Otani et al. 1996;
 Guainazzi et al. 1996) and thus, through the known ionization parameter (see
below) on the location of the clouds. 
 The external pressure difference
$\Delta P_{\rm ext} \equiv P_{\rm ext}(R)-P_{\rm ext}(R+\Delta R)$
 satisfies
$\Delta P_{\rm ext} /P_{\rm ext}= -\alpha \Delta R/R \ll  1$. 
This assumption is checked for self-consistency in \S3.5.
\item
The clouds are assumed to be in
photo-ionization equilibrium. This is justified as long
as the recombination time is the shortest time scale.
 This is true in our parameter range provided the ionizing
continuum is not rapidly varying (Krolik \& Kriss 1995). 
\item
The dynamical problem is treated by assuming radiation
pressure force, gas pressure force, and gravity.
The local gas acceleration, $a(r)$, at position $r$ inside a cloud
whose illuminated face is at a distance $R$ from the centre, is given by
\begin{equation}
a(r)=a_{\rm rad}(r)-g - \frac{1}{\rho} \frac{dP_{\rm gas}}{dr}
\label{accel}
\end{equation}
where $a_{\rm rad}$ is radiation pressure  acceleration 
pressure, $g$ the gravitational acceleration, $\rho$ the gas density, 
and $P_{\rm gas}$  the gas pressure. Note that
the dependence of $g$ and $a_{\rm rad}$ on $R$
has been omitted since $R$ is constant across the
geometrically thin cloud. Furthermore, we
assume quasi-hydrostatic cloud structure in the cloud's rest frame.
 As shown by Blumenthal \&
Mathews (1979), and others, this is an adequate description of 
 BLR and similar clouds. The clouds considered here are illuminated by the same
radiation field and are roughly at the same location, hence the same general
arguments apply. The clouds are therefore regarded as moving with  a
certain, distance dependent, bulk velocity and acceleration ($\left <
  a \right >$, i.e., $a(r)=\left < a \right >$),  with no significant velocity gradients. Given this,   
we do not have to treat the full hydrodynamic problem but assume that the
clouds remain stable and do not disperse as they are accelerated. The
quasi-hydrostatic assumption is checked for self-consistency in \S3.5.
\end{enumerate}

\subsection{Ionization structure and radiation pressure force}

The ionization structure of the cloud is determined by the local
radiation field. This is usually described in terms of the illuminated
face ($r=0$) ionization parameter, $U$,
\begin{equation}
U=\frac{\int_{\nu_0}^{\nu_1} \left (L_\nu/h\nu \right ) d\nu}{4\pi
  R^2 n_H c} 
\label{ionf}
\end{equation}
where $L_\nu$ is the monochromatic continuum luminosity
($\rmn{erg~s}^{-1}~\rmn{Hz}^{-1}$),  
$n_{H}$ the hydrogen number density at $r=0$ and $c$ 
the speed of light. Our aim is to study the acceleration of highly
ionized clouds. We therefore use the X-ray ionization parameter,
$U_x$ (Netzer 1996), defined by $h\nu_0=0.1$~keV and
$h\nu_1=10$~keV. 
The ionization and thermal structure are calculated by {\sc ion99}, the 1999
version of the photo-ionization code {\sc ion} (Netzer 1996, and references
therein). The code is very detailed regarding X-ray related processes
and solves for  the temperature profile and the radiation
pressure force distribution inside the cloud. All our calculations apply
to fully exposed clouds 
(i.e., no obscuration or shielding).

\begin{figure}
\centerline{\epsfxsize=3in\epsfbox{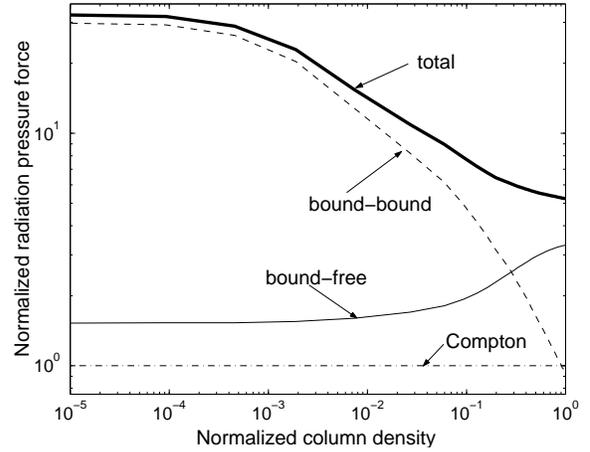}}
\caption{Contribution of various radiative processes to the normalized
  radiation pressure force (relative to the Compton radiation pressure
  force) for the standard case with line width given by
  $v_{\rmn{sound}}$. Note the sharp
  decline of the bound-bound contribution, after $\sim 1\%$ of the
  total column, due to optical depth in the lines.}
\label{force_m}
\end{figure}

 The absorption of continuum radiation by bound-bound, bound-free,
 Compton scattering and free-free processes results in a net momentum
 transfer to the clouds. The relative importance of the various
 processes depends on the atomic cross sections and the spectral energy
 distribution (SED). The largest cross sections are for bound-bound
 transitions. These occur at energies lower than the ionization
 energy of the relevant ion where, for the assumed  AGN-type SED   (see below)
the value of $L_\nu$ is  larger than its value at the ionization threshold.
However, large cross sections imply large optical depths and  self shielding.
Thus, bound-bound absorption contributes the most at  the illuminated surface
and decays quickly into the cloud  (Fig.~\ref{force_m}). Bound-free processes
are the next largest  contributers.  The Compton cross-section is relatively
small, and the  optical depth is negligible in all cases considered here
(column  density $<1.5 \times 10^{24}~\rmn{cm}^{-2}$). Thus, Compton scattering
 contributes equally (per particle) across the cloud. In almost all
 cases (see however \S3.1),   free-free absorption contributes very little
 to the radiation pressure force ($\sim 0.01$
 Compton). We neglect 
internal line radiation pressure (see Elitzur \& Ferland 1986) which
do not contribute significantly to the total pressure because of the
low excitation temperature of the strongest X-ray transitions.

We use the force multiplier formalism 
(e.g., Mihalas \& Mihalas 1999). The force
multiplier is defined as the ratio between the 
radiation pressure force due to all processes relative to the force due to
Compton scattering,
\begin{equation}
M(r)\equiv \frac{\int_0^{\infty}\chi_{\nu}(r)L_{\nu}e^{-\tau_\nu(r)}d\nu}{n_e(r)
  \sigma_T L_{\rmn{tot}}},
\label{m_def}
\end{equation}
where $\chi_\nu(r)$ is the frequency dependent total absorption and
scattering coefficient, 
$n_e$ is the number density of free electrons, $L_{\rmn{tot}}$
is the total, unattenuated ($r=0$) luminosity, and $\tau_\nu(r)$ is the
position dependent optical depth. This gives
\begin{equation}
a_{\rmn{rad}}(r)=\frac{n_e(r) \sigma_T L_{\rmn{tot}}}{4 \pi R^2 \rho(r) c} M(r). 
\label{frad}
\end{equation}

\subsection{Model parameters}

Several  parameters influence the solution.  
The SED detrmines the ionization
structure and the radiation pressure force over the various energy
bands. We have experimented with several SEDs typical of type-I AGN and
 found only little differences regarding the
structure and dynamics of the clouds. We therefore take as our
standard and only case an SED characterized by a ``UV bump'',$\alpha_{ox}=1.4$,
$\alpha_{(0.5-1~keV)}=1.06$, and $~\alpha_{(1-10~keV)}=0.9$.

The internal line widths determine the line optical depth  and thus the
 bound-bound contributions to \mav. We have
experimented with three possibilities: a) Thermal line width, $v_T$. b)
Gaussian-shaped profiles with  line widths for all ions equal to the thermal
hydrogen line width (denoted by $v_{\rmn{sound}}$). c)  Gaussian-shaped,
line  profiles, characterized by
their FWHM (typically several hundred ${\rm km~s^{-1}}$) and motivated by the Crenshaw et
al. (1999) and Kaspi et al. (2001) observations.  Mechanisms responsible for
such  broadening  have been discussed, extensively, in the
literature (see Bottorff \& Ferland (2000) for discussion of MHD turbulences,
including many older references, as well as  many papers discussing
microturbulences and expansion in stellar atmospheres). 

The range of ionization parameter and column density is determined by the
observations 
 (Reynolds \& Fabian 1995; Reynolds 1997; George et al. 1998; Kaspi et
 al. 2001). We chose
 $0.02<U_x<4$ and $10^{20} \leq N_H \leq
10^{23}~\rmn{cm}^{-2}$.  Many of the 
results pertain to a ``standard model'' defined by
$U_x=10^{-0.8},~n_H=10^8~\rmn{cm}^{-3},~N_H=10^{22}~\rmn{cm}^{-2}$, with line
width given by $v_T$. 
We consider only the following
solar-like composition, 
${\rm H:He:C:N:O:Ne:}$ ${\rm
Mg:Si:S:Ar:Fe}=10^4:10^3:3.7:1.1:8:1.1:0.35:0.37:0.16:0.03:0.4$

\subsection{Method of solution}

We follow the general method
described in Weymann (1976), Mathews \& Blumenthal (1977) and
Blumenthal \& Mathews (1979) where the 
cloud structure obeys the hydrostatic equation 
\begin{equation}
\frac{dP_{\rm gas}(r)}{dr}=\rho (r) \left [ a_{\rm rad}(r) -g -\left < a
\right > \right ],
\label{dpdr}
\end{equation}
where  $\left < a \right >$ is the average (bulk) acceleration. This
 is satisfied for every distance $R$.
The radiation pressure force is changing across the cloud due to
optical depth effects. This results in a non-uniform gas pressure
profile which must obey the boundary condition
$P_{\rm{gas}}(r=0)=P_{\rm ext}(R) \approx P_{\rm ext}(R+\Delta
R)=P_{\rm{gas}}(r= \Delta R)$. The  exact distribution depends on 
 the ionization structure, the level population of the
various ions, and the temperature at every point inside the cloud.

The cloud accelerates ballistically and we make no use of the
 continuity
equation. Nevertheless, the cloud column
density can change due to the  external conditions since it can 
 accrete mass, lose mass, or expand along the way. In this work we treat
 the 1D problem and consider the following $N_H(R)$:
1) constant mass gas shells which subtend a
constant solid angle, $\Omega \leq 4\pi$, i.e. $\Omega R^2 N_H={\rm
  const}$,  2) constant column 
density clouds. The latter model is
included as an example for extreme cases where the column density
decreases with distance much slower than $R^{-2}$. For example,
spherical constant mass clouds imply 
$N_H \propto R^{2/3\alpha}$. This allows us to study the
sensitivity of our results to different column densities. In
\S3.5 we consider several other implications of these assumptions.

To calculate the acceleration 
$\left < a \right >$, we integrate equation (\ref{dpdr}) with the
given boundary conditions (i.e., pressure) at both cloud surfaces. Given our
parameter space and set of assumptions, the contribution of the small
external pressure gradients to the acceleration via $\Delta P_{\rm
  ext}/(\rho \Delta R)$  is negligible ($\sim 10^{-4} g$)
and the bulk acceleration is given by   
\begin{equation}
\left < a \right >=\frac{\int_0^{\Delta R}\rho(r) a_{\rm{rad}}(r)
  dr}{\int_0^{\Delta R}\rho(r)dr} -g,
\label{aav}
\end{equation}
which depends on $R$ via the boundary conditions, the value of $N_H$
and  via $a_{\rm rad}(r)$. 

Keeping with the definition of the average acceleration, we  define the
average force multiplier,
\begin{equation}
\left< M \right > =\frac{\int_0^{\Delta R} n_e
  (r) M(r)dr}{\int_0^{\Delta R} n_e (r) dr},
\label{m_av_def}
\end{equation} 
where we have made use of the fact that  $U_x$ is large and hydrogen and
helium are fully ionized, thus
$\rho(r)/n_e(r) \sim 1.18m_H$.  
Thus we obtain the hydrostatic equation in the force multiplier representation,
\begin{equation}
\frac{dP_{\rm gas}(r)}{dr}=\frac{n_e(r) \sigma_T L_{\rmn{tot}}}{4
  \pi R^2 c} \left [ M(r)- \left <M \right > \right ].
\label{dpdrm}
\end{equation}

The cloud dynamics is therefore reduced to the ballistic case where
the cloud is accelerated coherently with the following equation of
motion 
\begin{equation}
v\frac{dv}{dR}=\left< a(R) \right >=\frac{1}{1.18m_H} \frac{ \sigma_T
  L_{\rmn{tot}}}{4 \pi c  
  R^2} \left< M (R) \right > -\frac{GM_{\rm BH}}{R^2}
\label{eqnbasic}
\end{equation}
where $v$ is the velocity of the cloud and $M_{{\rm BH}}$ is the mass of the
central black hole. Thus, $\left< M \right >$ is the only parameter
which depends on the inner structure of the cloud.  

Defining a new variable, $l$, which is proportional to $1/M_{\rm BH}$,
by 
\begin{equation}
l=\frac{1}{1.18m_H} \frac{ \sigma_T L_{\rmn{tot}}/c}{4\pi G
  M_{\rm BH}}=\frac{L_{\rmn{tot}}}{1.18L_{\rmn{edd}}} \, , 
\label{lratio}
\end{equation}
 simplifies the equation of motion to
\begin{equation}
v\frac{dv}{dR}= \frac{1}{1.18m_H} \frac{ \sigma_T L_{\rmn{tot}}}{4 \pi c  R^2}
\left (\left< M(R) \right > -\frac{1}{l} \right) \, .
\label{eqn_motion}
\end{equation}
In the above formalism, the dynamical problem is decoupled from the cloud
structure problem (i.e. the structure is independent of the cloud velocity).

The actual calculations start  by
assuming a uniform density profile and by solving for the 
temperature and radiation pressure force profiles. We calculate
\mav, solve
 the hydrostatic equation and obtain the first estimate of
$P_{\rm{gas}}(r)$. Given the first temperature profile, we obtain a new density
profile which is used, as an input, for the next iteration. This
procedure is repeated several times until convergence. Tests show that
 the final results
 depend very little on the initial  density profile. The iteration stops when
the density profile does not
 change by more than 1\% between one iteration to the next.

Given the pressure and ionization profile, the solution of the equation of
motion is straight forward. The initial conditions  include the
distance from the centre at the base of the flow, $R_0$, $P_{\rm
  ext}(R_0)$, $N_H(R_0)$, 
the column density as a function of $R$ (i.e., constant column clouds
or constant mass shells), and the AGN parameters ($\alpha$, $L_{\rm
  tot}$ and $l$). This allows us to solve for the cloud structure and
for $v(R)$ at each $R$.

\section{Results and Discussion}

This section describes the results of the calculations 
regarding the cloud's internal structure, the average force
multiplier $\left< M \right >$, and the cloud 
motion. Unless otherwise
specified the standard model parameters are assumed.

\subsection{Cloud structure}

\begin{figure}
\centerline{\epsfxsize=3in\epsfbox{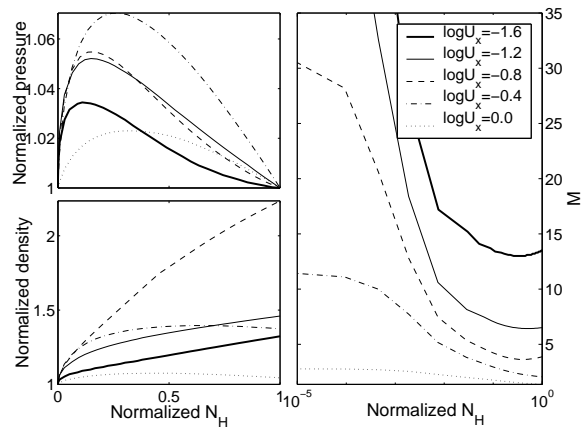}}
\caption{Density, pressure and force multiplier across the cloud for
  various values of $U_x$. In all cases $N_H=10^{22}~\rmn{cm}^{-2}$ and
the lines have thermal profiles.
The most
  pronounced difference is in $M(r)$, where for lower $U_x$,
  bound-bound processes contribute more near the illuminated surface
  with a sharper decline towards the outer surface.}
\label{struct_vs_U}
\end{figure}

We allow the ionization parameter to vary in the range
$10^{-1.8} - 10^{0.6}$ assuming all other parameters remain
constant. The contributions of the various absorption and scattering
processes to $M$ are computed from the ionization
structure. As seen in Fig. \ref{struct_vs_U}, the force multiplier
decreases with $U_x$. The reasons are the fewer bound-bound
transitions at large $U_x$ and the lower available driving flux at
high energies. 

A common property of all models is the increasing density into the cloud
(Fig.~\ref{struct_vs_U}), due to the temperature decrease as a
function of $r$ and the constant external pressure.
  The most extreme case shown here is  for $U_x=10^{-0.8}$ where the density 
 increases by a factor $\sim 2$ above its initial value. This is also
 the ionization parameter where the ionization and thermal structure
 are dominated by H-like and He-like oxygen.

Fig. ~\ref{struct_vs_U} also shows the internal pressure profile.
The rise near $r=0$ is caused by the pressure
gradient that must balance the radiation pressure force  which is larger at
this location due to the  contribution of bound-bound
processes. The peak pressure depends on the difference $M(r=0)-\left
  <M \right>$ and the location $r$ where $M(r)=\left <M
\right>$. For very large values of $U_x$, the peak is lower since the
gas is highly ionized and  bound-bound transitions are less important.

Finally, since the value of the temperature at the illuminated
surface, $T$, is independent of the column density, we can approximate
it by $T \simeq 10^{5.5} U_x^{0.5}$ which is a good approximation for
the abundances, SED and the range of $U_x$ considered here. The same
functional form, with different coefficients, provides a good fit for
other SEDs (see, e.g., Komossa \& Meerschweinchen 2000).

\begin{figure}
\centerline{\epsfxsize=3in\epsfbox{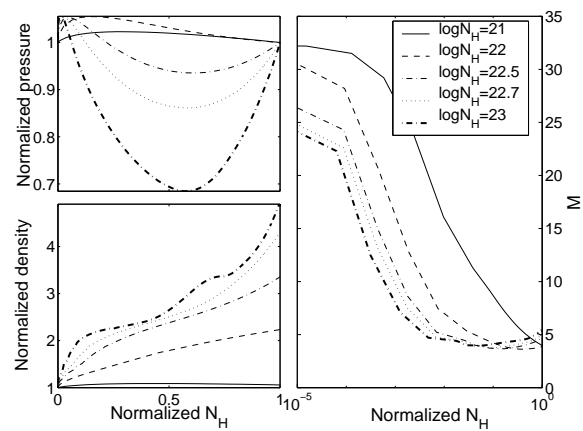}}
\caption{Cloud structure for various $N_H$. The cloud becomes
  highly non-uniform with increasing $N_H$ and the radiation pressure
  force rises towards the leading, non-illuminated edge (see the
  $N_H=10^{23}~\rmn{cm}^{-2}$ curve)}. 
\label{struct_vs_colu}
\end{figure}

Changes in column density affect  the
pressure and temperature profile because of the increased opacity. 
This is shown in Fig.~\ref{struct_vs_colu}. For low
column densities, and the standard $U_x$, the clouds have almost
uniform ionization structure. The ionization structure and the
radiation pressure force distribution  change
considerably for $N_H \geq 10^{22}~\rmn{cm}^{-2}$ since this column density
corresponds to an optical depth of $\sim 1$ in the \ovii\ and
\oviii\ bound-free transitions. 

A common feature for all column densities is the fast
decline of the bound-bound force multiplier, $M_{\rmn{bb}}$,  beyond $N_H \geq
10^{17}\rmn{cm}^{-2}$, due to the large line opacity.
 Deeper layers contribute much more to the bound-free force multiplier,
$M_{\rmn{bf}}$. This behavior is column density dependent since, for large columns,
the increased bound-free opacity causes lower energy bands of the continuum to
become more significant in
producing the local radiation pressure (for the typical SED considered
here, there are many
more photons in these bands). The increase in $M_{\rmn{bf}}$ is the reason for the
 peculiar pressure profile in the $N_H >10^{22}~\rmn{cm}^{-2}$ models.  

\begin{figure}
\centerline{\epsfxsize=3in\epsfbox{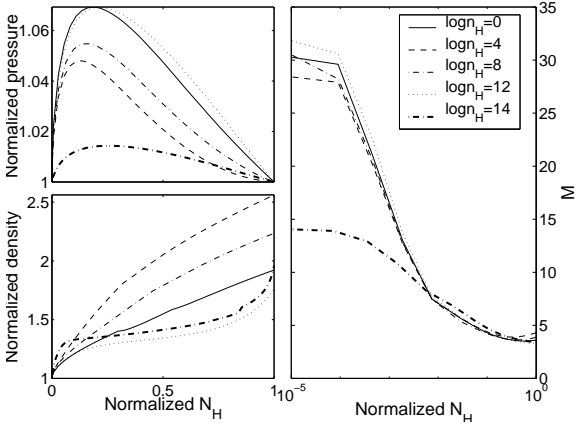}}
\caption{Density and pressure profiles for different values of
  $n_H(r=0)$. The shape of the density profile is similar in all cases
  up to $n_H \simeq 10^{12}~\rmn{cm}^{-3}$, beyond which the micro-physics
  change considerably (see the
  $n_H=10^{14}~\rmn{cm}^{-3}$ case  on the right-hand side).}
\label{struct_vs_n}
\end{figure}

We study  clouds with a large range of densities
($1<n_H<10^{14}~\rmn{cm}^{-3}$) at the illuminated surface. As shown in
Fig. \ref{struct_vs_n}, the cloud structure is not very sensitive to
$n_H$ for $n_H(r=0) \le 10^{11}~\rmn{cm}^{-3}$ since the micro-physics
remains essentially unchanged. All  changes can
therefore be reduced to a scale change in $\Delta
R$. Above $n_H=10^{11}~\rmn{cm}^{-3}$, the micro-physics changes
considerably due to  collisional excitation and free-free heating.
This results in  more
uniform temperature and density profiles and hence changes in $M_{\rmn{bb}}$.
In general, the very high density clouds are characterized by higher
  ionization, higher temperature and smaller $M$.

\begin{figure}
\centerline{\epsfxsize=3in\epsfbox{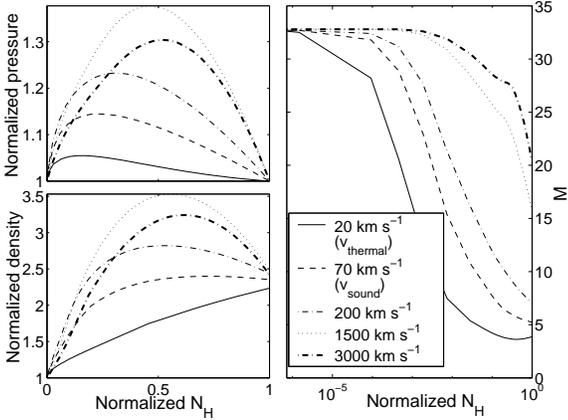}}
\caption{Cloud structure for several FWHM values of oxygen lines. The monotonic
  change in density, pressure, and $M(r)$ is the result of the
  increasing importance of bound-bound transitions. Note that for
  FWHM=3000 $\rmn{km~s}^{-1}$, the bound-bound processes contribute
  significantly to $M(r)$ even at the leading (non-illuminated)
  surface.}
\label{struct_vs_v}
\end{figure}

The internal velocity spread is dynamically important.
Wider profiles decrease the line opacity and increase
the thickness of the zone where $M(r)$ is dominated by bound-bound
transitions. This shifts the peak of the
pressure profile deeper into the cloud (Fig.~\ref{struct_vs_v}). 
A large FWHM does little to affect the
 local ionization and temperature balance, as expected from purely
radiative considerations. However, the increased importance of
$M_{\rmn{bb}}$ results, in low column
density clouds, in a  more uniform pressure profile.

\subsection{The dependence of \mav~ on $U_x$, $N_H$,
  $n_H$ and the line width} 

The average force multiplier is the primary  property determining 
the cloud velocity. It is therefore important to investigate its behavior
as a function of the cloud properties.

\begin{figure}
\centerline{\epsfxsize=3in\epsfbox{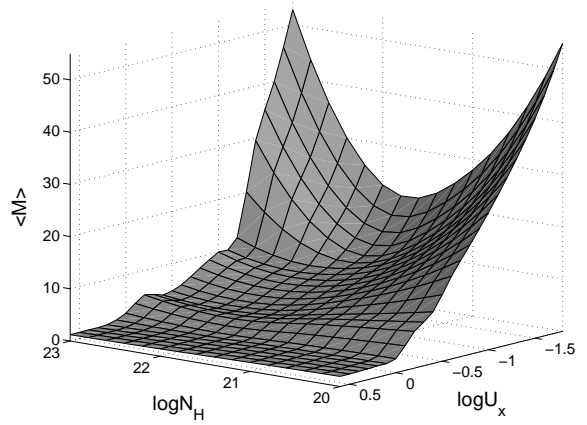}}
\caption{$\left< M \right >$ as a function of the ionization
  parameter, $U_x$, and the column
  density, $N_H$. A monotonic decline of $\left< M \right >$ with $U_x$ is
  evident for all  $N_H$ . For $U_x>10^{0.5}$,
  the cloud is highly ionized and Compton scattering dominates the
  radiation pressure force. $\left< M \right >$ obtains a
  minimum as a function of $N_H$ for $N_H \sim
  10^{22}~\rmn{cm}^{-2}$ for $U_x<1$. The increase in $\left< M \right
  >$ , for $N_H>10^{22}~\rmn{cm}^{-2}$, is the result of the changing
  ionization balance (see text).}
\label{m_av_vs_ucolu}
\end{figure}

The dependence of  $\left< M \right >$ on the ionization parameter,
$U_x$ is shown in
Fig. \ref{m_av_vs_ucolu}. Since the clouds are more ionized for
larger $U_x$,  $\left< M \right >$ is more sensitive to
higher frequency bands, where the incident flux is lower. The result
is that $\left< M \right >$ is a decreasing function of $U_x$
 for all column densities. 
For low column
densities ($N_H<10^{22}~\rmn{cm}^{-2}$) the change in $\left< M \right >$ is
roughly linear with $U_x$. For larger
columns,  the  structure is more complicated
(e.g., Fig.~\ref{struct_vs_colu}) and so is the dependence on $U_x$.
A crude  approximation  of
the form ${\rm log}\left< M(U_x) \right > = C_1+C_2{\rm log}(U_x)$, for several
column densities and for line widths given by $v_{\rmn{sound}}$, is
\begin{equation}
{\rm log}\left ( \begin{array}{c}
        \left < M \right >_{20} \\
        \left < M \right >_{21} \\
        \left < M \right >_{22} \\
        \left < M \right >_{23}
\end{array} \right )= 
\left ( \begin{array}{c}
        -0.7 \\
        -0.6 \\
        -0.5\\
        -0.7
\end{array} \right ) {\rm log}(U_x)+
\left ( \begin{array}{c}
        0.6 \\
        0.5 \\
        0.4 \\
        0.4
\end{array} \right ),
\label{mavfit}
\end{equation}
where $\left < M \right >_n$ is the average force multiplier for
$N_H=10^n~\rmn{cm}^{-2}$. This approximates $\left< M \right >$ within a
factor of $\sim 2$. 

Fig. \ref{m_av_vs_ucolu} also shows the dependence of $\left< M \right >$ on the
column density. This is characterized by a decrease followed by an increase
with a minimum 
at $N_H \sim 10^{22}~\rmn{cm}^{-2}$.  The minimum is mostly due to the fact that at
this column density, the optical depth due to absorption by oxygen is close to unity.
For $N_H \leq
10^{22}~\rmn{cm}^{-2}$, the clouds are ionized throughout  
(Fig.~\ref{struct_vs_colu}) and the decline in $\left< M \right >$ reflects a 
decrease in the available X-ray flux. 
For $N_H>10^{22}~\rmn{cm}^{-2}$, the ionization structure is such that the leading edge is
neutral enough to absorb a large fraction of the far UV
radiation. Since the photon flux 
at those bands is much larger, it results in an increasing $\left< M \right >$.
 The changes in $\left< M \right >$
 are more pronounced for low values of $U_x$ because of the SED shape.

\begin{figure}
\centerline{\epsfxsize=3in\epsfbox{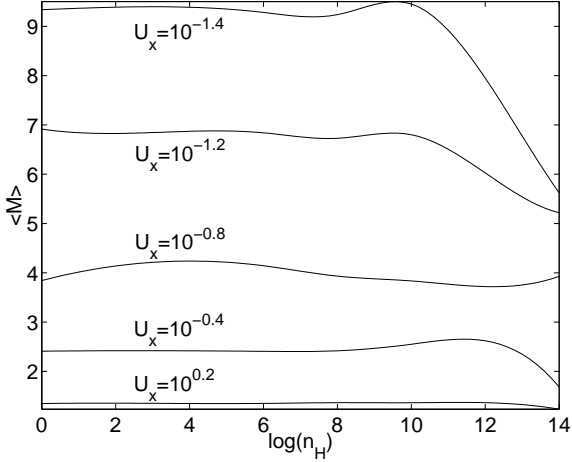}}
\caption{Average force multiplier as a function of $n_H(r=0)$ for the
  standard model with $v_T$ and different $U_x$. For $n_H<10^{11}~\rmn{cm}^{-3}$,
  $\left< M \right >$ is roughly constant. The changes in $\left< M
  \right >$ at high densities are due to the changes in the
  micro-physics of the gas.} 
\label{m_av_vs_nh}
\end{figure}

The gas density  has only a small influence on
the cloud structure, and on $\left< M \right >$, for
$n_H(r=0)<10^{11}~\rmn{cm}^{-3}$ (Fig.~\ref{m_av_vs_nh}). 
 For
higher densities, the micro-physics changes (see \S3.1) and the changes
in $\left< M \right >$ can amount to a factor $\sim2$ between
$n_H(r=0)\sim 10^8~\rmn{cm}^{-3}$ and $n_H(r=0)\sim 10^{14}~\rmn{cm}^{-3}$.
 As shown below, for the $P_{\rmn{ext}} \propto
R^{-2}$, $N_H=\rmn{const}.$ case, this corresponds to a change of $\sim40\%$ in 
terminal velocity. Such changes can be critical for
marginally accelerating clouds.

\begin{figure}
\centerline{\epsfxsize=3in\epsfbox{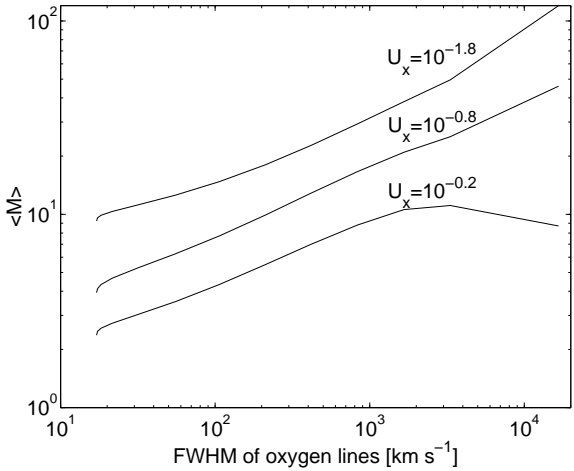}}
\caption{$<M>$ as a function of the FWHM of oxygen lines. Note the
  logarithmic scales, and the incline of $\left < M \right>$ (for
  $U_x<10^{-0.2}$) towards large FWHM values, which characterise BAL flows.}
\label{m_av_vs_v}
\end{figure}

The dependence of \mav~  on the FWHM of oxygen lines is shown in
Fig. \ref{m_av_vs_v} where we consider
FWHM$<16,000~\rmn{km~s}^{-1}$, similar to BAL flow velocities. We do not
show cases with $v_{\rmn{sound}}$ which 
are roughly equivalent to FWHM$\sim 70~\rmn{km~s}^{-1}$.  
As expected, broader lines produce larger force multipliers ($\sim 6$
times larger for FWHM$\sim 3000~\rmn{km~s}^{-1}$ compared with thermal case). The
effect is more pronounced for small $U_x$ where there
are more lines available to absorb the larger continuum flux. For
 $U_x=10^{-0.2}$ there is a decrease in \mav~  for
 FWHM$\gtorder 3000~\rmn{km~s}^{-1}$ due to changes in ionization
 structure which mask the relatively small increase in line
 contributions. For FWHM$\gtorder 16,000~\rmn{km~s}^{-1}$, $\left < M
 \right >$ is relatively constant since the lines are optically
 thin relative to the bound-free processes.

While the cloud structure and, therefore, the radiation pressure force are
relatively insensitive to the exact 
shape of the SED (i.e., $\alpha_x$, $\alpha_{ox}$) for the same $U_x$, the exact value of $\left < M
\right >$ depends roughly linearly on the X-ray (0.1-1keV) to total
luminosity ratio (equation \ref{m_def}) which in our case is $\sim0.04$. 

\subsection{Approximations for \mav}

Fig. \ref{v_profs}a illustrates the most important
contributions to $\left< M \right >$ with respect to the cloud
dynamics. While the exact value depends on the
cloud's structure, the bound-free force multiplier, $M_{\rmn{bf}}$, is, in general, a
good approximation to  \mav. 
It is possible to estimate $M_{\rmn{bf}}$ by considering 
oxygen, the most abundant metal. Assuming cross sections
of the form  $\sigma=\sigma_0(E/E_0)^{-s}$, where $E_0$ is the
ionization energy, $\sigma_0$ the edge cross section, $s\simeq
2.5-3$, and an X-ray spectrum with an energy power-law index $\alpha_x$. For 
 an ion $X$ with a number density
$n_x$ we find
\begin{equation}
M_{\rmn{bf}}=
\frac{1}{\alpha_x+s-1} \left ( \frac{n_X}{n_e} \right ) \left
  ( \frac{\sigma_0}{\sigma_T} \right ) \left (
  \frac{L_E(E_0)E_0}{L_{\rmn{tot}}} \right ),
\label{mbf}
\end{equation}
where $L_E(E_0)$ is the monochromatic flux  at the ionization edge. For
the range of interest, $10^{-1.8} \le U_x \le 10^{-0.2}$, and the
chosen SED, the main
contributions to the force multiplier are \ovii, \oviii, and
\heii. Applying the relevant constants,  
and assuming $X_{\rmn{\ovii}}+X_{\rmn{\oviii}}\simeq 1$, we obtain the following
 estimate:
\begin{equation}
M_{\rmn{bf}} \simeq  1.6+1.3X_{\rmn{\oviii}}+2\left ( \frac{X_{\rmn \heii}}{10^{-4}} \right ),
\label{mval}
\end{equation}
where $10^{-6} \ltorder X_{\rmn{\heii}} \ltorder 10^{-4}$. This  $M_{\rmn{bf}}$ is within a
factor $\sim 3$ of \mav~  for line width given by
$v_T$. This result as well as the general (all elements) $M_{\rmn{bf}}$ is
{\it significantly} different from the force multiplier calculated by
Mathews \& Veilleux (1989) who considered gas with much lower
ionization stages and calculated only the contribution due to the carbon
edge. For the typical $U_x$ and SED
considered here this gives  a value of \mav ~which is a factor of
50(!) smaller than the one calculated by us.

\subsection{Cloud dynamics}

Solving the equation of motion (equation \ref{eqn_motion}) requires
the value of 
 $\left < M(R) \right>$ for
every $R$. The formal solution, assuming $v(R_0)=0$, is
\begin{equation}
v(\xi)=v_c \left [ \int_{1}^{\xi} \left ( \left < M(\xi') \right >
    -\frac{1}{l} \right ) \frac{d\xi'}{\xi'^2} \right ]^{1/2},
\label{formals}
\end{equation}
where $\xi=R/R_0$. The characteristic velocity $v_c$ is  defined as
\begin{equation}
v_c=\sqrt{ \frac{1}{1.18m_H} \frac{ \sigma_T L_{\rmn{tot}}}{2\pi c R_0}} \simeq
1350\left ( \frac{L_{45}}{R_{17}} \right )^{1/2}\rmn{km~s}^{-1},
\label{vc}
\end{equation}
where $L_{45}$ and $R_{17}$ are the luminosity in units of
$10^{45}~\rmn{erg~s}^{-1}$ and the initial distance ($R_0$) in units of
$10^{17}~\rmn{cm}$. The dynamical problem is thus reduced to the calculation
of a single 
quantity, $\left < M(R) \right >$, at each position along the cloud
trajectory. 

\begin{figure}
\centerline{\epsfxsize=3in\epsfbox{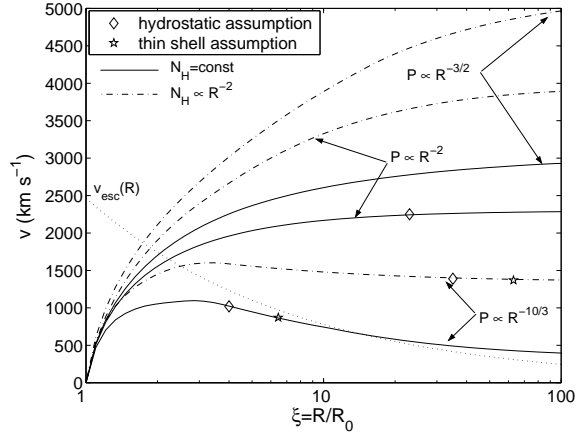}}
\caption{ The dependence of $v(\xi)$ on $P_{\rmn{ext}}$ for a case of 
  $U_x=10^{-0.8},~N_H(R_0)=10^{22}~\rmn{cm}^{-2},~R_0=10^{17}~\rmn{cm},
  ~L=10^{45}~\rmn{erg~s}^{-1},~l=0.3$. Constant $N_H$ cloud profiles in solid lines,
  and constant mass shell ($N_H\propto R^{-2}$) in dashed line. 
  The escape velocity, $v_{\rmn{esc}}(R)$, is
  shown for comparison. For $\alpha=-10/3$, gravity takes over due to
  the rapid expansion of the cloud. Note the critical points beyond which the
  hydrostatic and the thin shell assumptions are no longer valid (see text).}
\label{v_profs2}
\end{figure}

We have studied  velocity profiles for two generic
cases: shell-like clouds with
 constant mass ($\Omega R^2N_H=\rmn{const}$) and 
constant column density clouds ($N_H(R)=\rmn{const}$). The exact numerical
solutions are shown in Fig. \ref{v_profs2}. The diagram illustrates the
 lower velocities that
are obtained  for the constant column density clouds
since these are more massive  (note that 
 a comparison between the two is not straight forward since 
 \mav~  is a non-monotonic function of $N_H$, e.g., \S3.3).
For the standard model, the velocities are of the same order as the escape
velocity at the base of the flow. In particular, velocities of
1000--3000 $\rmn{km~s}^{-1}$ are typical of HIG clouds originating in or just
outside the BLR. Much larger velocities are achieved for
higher density clouds that originate inside the BLR (related, perhaps,
to the inner accretion disc). For example, the standard model
 with $n_H=10^{12}~\rmn{cm}^{-3}$ and constant column density
 results in a terminal velocity of $\sim 20,000 ~\rmn{km~s}^{-1}$ for line
 width given by $v_{\rmn{sound}}$.  
Thus, acceleration by X-ray radiation pressure is capable of producing
BAL-type flows in AGN. This conclusion applies only to the ballistic motion of
hydrostatically stable clouds. Wind-type flows may be different and
will be discussed in future works.

Three cases of external pressure
powerlaws were examined for constant column density clouds and for
constant mass shells, all with
$n_H <  10^{11}~{\rm {cm}^{-3}}$ (see Fig. \ref{v_profs2}). The main
results are explained using the constant column density cloud case.

\begin{figure}
\centerline{\epsfxsize=3in\epsfbox{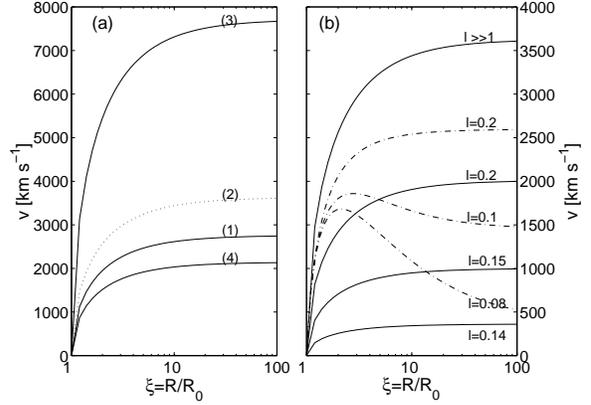}}
\caption{Velocity profiles for
  $U_x=10^{-0.8},~N_H=10^{22}~\rmn{cm}^{-2},~R_0=10^{17}~\rmn{cm},~L=10^{45}~\rmn{erg~s}^{-1},~l \gg 1$,  
  and $P_{\rmn{ext}} \propto R^{-2}$. Left: $v(\xi)$ for different 
  approximations of $\left< M \right >$: (1) Exact solution for
  thermal line width ($v_T$). (2) like (1) but line width is given by
  $v_{\rmn{sound}}$, (3) $\left< M \right >=M(r=0)$, 
  (4) $\left< M \right >=M_{\rmn{bf}}(r=0)$. As illustrated, the cloud's
  dynamics is dominated by 
  bound-free absorption for both $v_T$ and $v_{\rmn{sound}}$ line profiles. 
  Right: $v(\xi)$ for different values of $l$. line width is given by
  $v_{\rmn{sound}}$. Note the large differences in $v_{\rmn{final}}$ for $1/l \simeq
  \left < M \right >$. Solid lines are for clouds with pure radial
  motion. Dash-dot lines are for clouds with asimuthal (Keplerian)
  velocities at $\xi=1$ (see text). Super Eddington luminosities are
  included for completion.}
\label{v_profs}
\end{figure}

The first case $U_x(R)=\rmn{const}$ ($\alpha=-2$).  Here 
 $\left < M(R) \right >$ is a
constant of motion and equation (\ref{formals}) results in 
\begin{equation}
v(R)=v_{\rmn{final}} \sqrt{1-\frac{R_0}{R}}.
\label{scaled_v_prof}
\end{equation}
where the final (asymptotic) velocity is
\begin{equation}
v_{\rmn{final}}= v_c \left [ \left < M(U_x) \right > -\frac{1}{l}
\right ]^{1/2}.
\label{v_final}
\end{equation}
The final velocity of marginal outflows is therefore very sensitive to
the value of $l$ since $1/l \simeq \left < M \right >$
 (Fig.~\ref{v_profs}b)    
The acceleration time-scale, defined as the time it takes the
cloud to reach 90\% of its asymptotic velocity, is 
\begin{equation}
t_{{\rm accel}} \simeq 150  \frac{v_c}{v_{\rmn{final}}} R_{17} ~ {\rm years}.
\label{tflow}
\end{equation}

In the two other cases, $U_x$ decreases with $R$ ($\alpha > -2$) 
and $U_x$ increases with $R$ ($\alpha < -2$) (see
Fig. \ref{v_profs2}). In the first case, the
acceleration due to the radiation pressure
force decreases as $1/\xi$ to some (positive) power for our range of
$\alpha$. For $\xi \gg 1$, the clouds become asymptotically 
neutral and the radiation acceleration drops to zero.

In  the second case, $U_x$ increases with $R$, the clouds become
more ionized as they move out, bound-bound and bound-free processes become
less efficient and at large $\xi$, the only driving force is due to
Compton scattering. 

The condition for  asymptotic outflow ($\xi \gg 1$) is that  
 the cloud velocity exceeds the escape velocity at some distance, 
\begin{equation}
v(R)>v_{\rmn{esc}}(R)=\sqrt{ \frac{1}{1.18m_H} \frac{\sigma_T L_{\rmn{tot}}}{2\pi  c R
    l}}=v_c\sqrt{\frac{1}{l}}. 
\label{outcon}
\end{equation}
Contrary to other wind  solutions, in which the final
velocity is of the order of $v_{\rmn{esc}}(R_0)$, the final velocity
in our model   
depends on  the value of $ \left [M(R) l-1 \right]$ (equation \ref{formals})
which can be much larger than unity. 

Fig. \ref{v_profs2} also shows that a fair fraction of the final velocity
is obtained at small $\xi$. This is the reason for the similar
terminal velocity in 
the  $\alpha=-2$ and the $\alpha=-3/2$ cases. 
A different behaviour is seen for $\alpha=-10/3$ where the
gravitational force exceeds the radiation pressure force at
$\xi \sim 3$, resulting in a marginal outflow.

The dynamics of constant mass gas shells follows the same trend
 for the various external pressure profiles, but with
higher terminal velocities. This results from the decrease in column
density with distance and the fact that for $N_H<10^{22}~{\rm
  cm}^{-2}$, \mav ~is a monotonically decreasing function of $N_H$ (see
Fig. \ref{m_av_vs_ucolu}).

So far we have considered radial motion only. Other AGN components,
such as the BLR and the
accretion disc, are thought to be rotating around the central mass,
and a plausible assumption is to associate a rotational motion to the
HIG. Introducing Keplerian velocities (see also \S 3.5) at $R_0$, while
conserving angular momentum along its trajectory, will
decrease the effective gravity (in equation ~\ref{eqn_motion}, the term $1/l$
will be replaced by $1/l(1-1/\xi')$) thus increasing the final
velocities (see Fig.~\ref{v_profs}b). 

The obtained dynamical results differ significantly from the few
previous calculations involving acceleration of clouds by X-ray
radiation pressure force. It is an order of
magnitude difference from the Mathews \& Capriotti (1985) results (see
their table 1) and this is not surprising since previous works did not
consider the relevant parameter space for the HIG and, therefore, did
not include the relevant radiation absorption mechanisms. 

\subsection{Limitations and extensions of the model}

We have made two critical assumptions: thin shell clouds (which is
always true, by construction, at $R_0$) and  hydrostatic equilibrium
at all locations. Below we derive general expressions for a critical,
normalized distance, $\xi_c$, beyond which one or both of these assumptions
fail.
   
The critical distance for the thin-shell approximation is
\begin{equation}
\xi_c^{({\rm thin})} = 
\left \{ 
\begin{array}{ll} 
\left ( \frac{R_0}{\Delta R_0} \right )^\frac{1}{\beta -\alpha-1} &
\beta-\alpha>1 \\
\infty & \rmn{otherwise}
\end{array}
\right .
\label{dyngood2}
\end{equation} 
where we have used $n_H \propto P_{\rm ext}$. Here, $\beta=0,-2$ for constant
column density clouds and constant mass shells, respectively. 
Since the initial conditions are such that 
 $\Delta R_0 /R_0 \ll 1$, we see that in general 
 $ \xi_c^{\rm (thin)} \gg 1$.

A quasi-hydrostatic cloud structure can
be obtained provided the cloud's sound crossing time is shorter than
the time it 
takes the external conditions (i.e., external pressure) to change
considerably (e.g., Blumenthal \& Mathews 1975) as a result of the cloud's
motion. This time is roughly  $R_0/v_{\rm final}$. The
quasi-hydrostatic approximation is therefore  valid provided $\xi \ll
\xi_c^{({\rm hyd})}$ where 
\begin{equation}
\xi_c^{({\rm hyd})} \sim \left (\frac{v_{\rm sound}}{v_{\rm
      final}}\frac{R_0}{\Delta R_0} 
\right ) ^\frac{1}{\beta - \alpha}.
\label{tsletd}
\end{equation}
In cases where $\beta-\alpha \leq 0$, the cloud structure and dynamics
can be treated by our model for all $\xi \geq 1$ provided $R_0v_{\rm
  sound}/(\Delta R_0 v_{\rm final})>1$. In conclusion, cases for which
 $\xi_c={\rm min}(\xi_c^{({\rm thin})},\xi_c^{({\rm hyd})}) <1$ cannot be treated by our model.

Considering specific examples, we note that for the standard model,
with
$L_{45}=1,~R_{17}=1,~l=0.3,~U_x(R_0)=10^{-0.8}$, and $\alpha=-2$, numerical
calculations give $\xi_c \simeq 20$. At $\xi=4$ the cloud is
already moving at 90\% of its final velocity. For two other cases of
 $\alpha=-2$ ($N_H \propto R^{-2}$) and $\alpha=-1.5$, the
approximations hold throughout the region. The case of
$\alpha=-10/3$ is different due to the rapid expansion of the
cloud. In this case, for $N_H={\rm const}$, our model approximations
break at small $\xi$ ($\xi_c \simeq 4$). 

Throughout this paper we assumed an external pressure that depends
only
on $R$. This is a natural assumption for the constant mass (thin
spherical shell) case. For the other case, where the column density
decrease
with distance is slower than $N_H \propto R^{-2}$, we can imagine a
situation where the cloud 
subtends a small
solid angle ($\Omega<4\pi$) and the external pressure profile is angle
dependent.
Clouds with finite lateral extents are prone
to evaporation through their rims due to pressure gradients inside the
cloud (e.g., Mathews \& Veilleux 1989, equation 1). 
This sets a lower limit on the cloud's lateral extents
and conversly on the spatial angle it subtends of
\begin{equation}
\frac{\Omega}{4\pi} \gtorder 10^{-2}\left ( \frac{v_{\rm sound}}{v_{\rm final}}
\right ) ^2.
\label{omega}
\end{equation}
Filamentary structure, and different dynamics in different directions, can be
 introduced into our formalism. The treatment of the
general two dimensional problem is beyond the
scope of this paper (for simple pancake shaped clouds see
Blumenthal \& Mathews 1979).

In this work we have considered general confinement, having in mind
external magnetic pressure.  Confinement by hot thermal gas is another
possibility that has been discussed, extensively, in the literature.
The presence of highly ionized, dilute gas between the clouds cannot
be ruled out. 
Relative velocities between such gas and the cloud will result in drag forces
that tend to decrease the final velocities. The cloud-gas
boundary may be subjected to various 
 instabilities which may affect the cloud
 structure considerably and ultimatly lead to its destruction. This
 complex situation cannot be solved using
the methods adopted here.

\section{Conclusions}
We have made detailed numerical calculations of the dynamics of highly
ionized gas clouds
that are in pressure equilibrium with external magnetic pressure and
are ionized by a typical AGN continuum. The self-consistent
calculations include the ionization and thermal equilibrium of the cloud, the
 radiative transfer and the
one-dimensional hydrostatical
equilibrium solution. The principal conclusion is that such highly
ionized clouds can be accelerated to high velocities by
means of radiation pressure acceleration. The
dynamical problem can be reduced to the calculation of a single
parameter, the average force multiplier, $\left < M \right>$. 
The 
terminal velocity of the flow scales like the escape velocity at its origin but
can exceed this velocity  by a large factor 
($\sim \sqrt{\left < M \right>}$). The cloud velocity is sensitive to the
confining pressure profile ($\alpha$), especially for marginal flows.
In
particular, flows  that originate just outside the BLR in objects with
 $L_{\rmn{tot}} \simeq 10^{45}~\rmn{erg~s}^{-1}$,
will reach velocities of $1000-3000~\rmn{km~s}^{-1}$, similar to the
velocities measured for the X-ray and UV absorption
lines. Highly ionized AGN clouds are driven mainly by
bound-free absorption and 
bound-bound processes are less important unless significant
non-thermal line broadening or very
low column densities ($<10^{20}~\rmn{cm}^{-2}$) are involved.

We thank Nahum Arav for useful discussions, and the referee
for some valuable comments. We acknowledge financial support by the Israel
Science Foundation and the Jack Adler Chair of Extragalactic Astronomy.

\end{document}